# D4M 3.0

Lauren Milechin[*], Alexander Chen[+], Vijay Gadepally[*+], Dylan Hutchison[$], Siddharth Samsi[*], Jeremy Kepner[*+]

[*]MIT Lincoln Laboratory, Lexington, MA, U.S.A.
[+]MIT CSAIL, Cambridge, MA, U.S.A.
[$]University of Washington, Seattle, WA, U.S.A.

## ABSTRACT

**Abstract** – The D4M tool is used by hundreds of researchers to perform complex analytics on unstructured data. Over the past few years, the D4M toolbox has evolved to support connectivity with a variety of database engines, graph analytics in the Apache Accumulo database, and an implementation using the Julia programming language. In this article, we describe some of our latest additions to the D4M toolbox and our upcoming D4M 3.0 release.

## I. INTRODUCTION

The D4M (Dynamic Distributed Dimensional Data Model) tool is an analytical library that allows flexible data representation and manipulation (Kepner, et al., 2012). D4M supports a variety of operations based on the mathematical structure of associative arrays. Associative Arrays can represent many types of data, including graphical, numeric, and string data, and supports a variety of arithmetic and set operations have a wide variety of uses (Huang, 2015) (Gadepally, 2015). Associative arrays are amenable to performing linear algebraic operations on heterogeneous data. Numerous recent additions to D4M have prompted us to release the next version: D4M 3.0. The first two additions expand D4M's database connectivity and computation capabilities, and the third addition is an implementation of D4M in the Julia programming language.

## II. D4M AND DATABASES

Part of D4M's capabilities include seamless interaction with a variety of database systems. Our previous work largely focused on support of the scalable key-value store Apache Accumulo. With the recent trends of "many sizes" and in-database analytics, we have extended the D4M connections to the relational databases PostGRES and MySQL and the array data store SciDB.

SciDB is a database designed for multidimensional, scientific data that uses an array data model and provides the ability to perform basic linear algebra operations on data within the database, without the need to query that data first (Stonebraker, Brown, Poliakov, & Raman, 2011). The D4M-SciDB connector allows a user to connect to SciDB, bind to an array, ingest data, and query data using familiar associative array syntax. For the purpose of D4M, SciDB arrays are nothing but associative arrays. The D4M-SciDB connector has demonstrated fast ingest, achieving peak ingest performance of nearly 3 million inserts per second (Samsi, et al., 2016). The D4M associative array model further allows for translation of data between Accumulo, SciDB and PostGRES. Within the BigDAWG polystore system, the D4M toolbox is currently used as the text island (Elmore, et al., 2015).

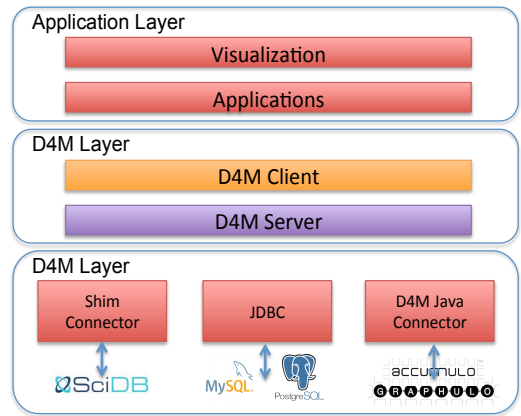

**Figure 1: D4M Architecture. D4M server bindings leverage various database connectors.**

The second addition to the D4M toolbox is the ability to perform fast in-database analytics in the NoSQL key-value store Apache Accumulo. D4M has been working seamlessly with the NoSQL database Accumulo for some time now, providing a schema (Kepner, et al., 2013), and a simple means to bind to tables and query and ingest data. Past work has shown record-breaking ingest performance using the D4M Schema and ingest (Kepner, et al., 2014). As part of the D4M 3.0 release, Graphulo, an extension on Accumulo, is also being released. Graphulo provides in-database graph operations outlined in the GraphBLAS, implemented as Accumulo server-side iterators. The D4M-Graphulo tool allows users to describe their graph analytics in the familiar GraphBLAS constructs and leverage the parallel infrastructure of Apache Accumulo to perform these operations directly in database without first transferring a partial set of results to local memory. These operations, such

This material is based upon work supported by the National Science Foundation under Grant No. DMS-1312831. Any opinions, findings, and conclusions or recommendations expressed in this material are those of the author(s) and do not necessarily reflect the views of the National Science Foundation.

as matrix multiply, enable many useful graph algorithms, including breadth first search, Jaccard coefficient, and k-truss. Graphulo have been shown to scale well to multi-node Accumulo instances (Weale, Gadepally, Hutchison, & Kepner, 2016) and outperform the client-side alternative in many cases. Our extensive performance results indicate that D4M-Graphulo can be used in cases where data size makes operations impossible to complete client-side due to memory constraints (Hutchison, Kepner, Gadepally, & Fuchs, 2015) (Hutchison, Kepner, Gadepally, & Howe, 2016). We have performed significant experiments with the D4M-Graphulo tool and have compared it to numerous parallel processing paradigms. As Figure 2 indicates, Graphulo can perform operations such as table multiplication at rates close to the in-memory D4M version without the same memory limitations.

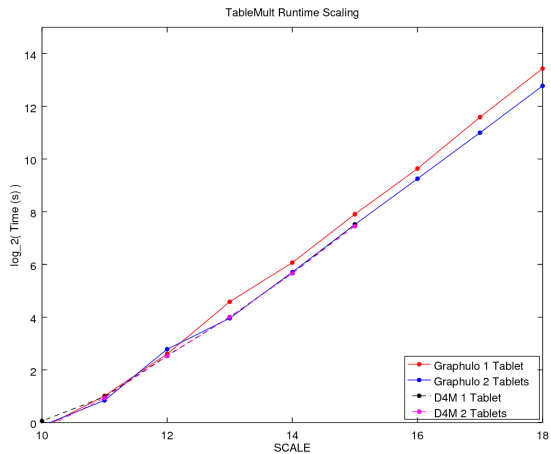

Figure 2: Graphulo vs. D4M TableMult Scaling

### III. D4M AND THE JULIA LANGUAGE

D4M can be implemented in any language that provides sparse linear algebra operations. Our previous
versions have been implemented in MATLAB® and required either MATLAB® or GNU Octave to use. The D4M.jl binding allows users to use D4M from the Julia programming language. Benchmarking work has shown D4M.jl operations to have comparable performance to MATLAB® D4M operations, and in some cases surpass previous operations (Chen, Edelman, Kepner, Gadepally, & Hutchison, 2016). The D4M.jl toolbox is open sourced and available for download.

### IV. CONCLUSIONS AND FUTURE WORK

The D4M 3.0 release incorporates many developments over the past 3 years. We believe that greater support for database engines, support for in-database operations, and wider language support are important additions to the D4M system.

### V. ACKNOWLEDGEMENTS

The authors wish to acknowledge the following individuals for their contributions: Michael Stonebraker, Sam Madden, Bill Howe, David Maier, Chris Hill, Alan Edelman, Charles Leiserson, Dave Martinez, Sterling Foster, Paul Burkhardt, Victor Roytburd, Bill Arcand, Bill Bergeron, David Bestor, Chansup Byun, Mike Houle, Matt Hubbell, Mike Jones, Anna Klein, Pete Michaleas, Julie Mullen, Andy Prout, Tony Rosa, and Chuck Yee.

### REFERENCES

Chen, A., Edelman, A., Kepner, J., Gadepally, V., & Hutchison, D. (2016). Julia Implementation of the Dynamic Distributed Dimensional Data Model. *IEEE High Performance Extreme Computing (HPEC)*. IEEE.

Elmore, A., Duggan, J., Stonebreaker, M., Balazinska, M., Cetintemel, U., Gadepally, V., et al. (2015). A Demonstration of the BigDAWG Polystore System. *Proceedings of the VLDB Endowment*. *8*, pp. 1908-1911. VLDB Endowment.

Gadepally, V. H. (2015). Computing on masked data to improve the security of big data. *Technologies for Homeland Security (HST), 2015 IEEE International Symposium*. IEEE.

Huang, Y. Y. (2015). A database-based distributed computation architecture with Accumulo and D4M: An application of eigensolver for large sparse matrix. *Big Data (Big Data), 2015 IEEE International Conference on*. IEEE.

Hutchison, D., Kepner, J., Gadepally, V., & Fuchs, A. (2015). Graphulo Implementation of Server-Side Sparse Matrix Multiply in the Accumulo Database. *IEEE High Performance Extreme Computing (HPEC)*. IEEE.

Hutchison, D., Kepner, J., Gadepally, V., & Howe, B. (2016). From NoSQL Accumulo to NewSQL Graphulo: Design and Utility of Graph Algorithms inside a BigTable Database . *IEEE High Performance Extreme Computing (HPEC)*. IEEE.

Kepner, J., Anderson, C., Arcand, W., Bestor, D., Bergeron, B., Byun, C., et al. (2013). D4M 2.0 schema: A general purpose high performance schema for the Accumulo database. *IEEE High Performance Extreme Computing Conference (HPEC)*. IEEE.

Kepner, J., Arcand, W., Bergeron, W., Bliss, N., Bond, R., Byun, C., et al. (2012). Dynamic Distributed Dimensional Data Model (D4M) Database and Computation System. *2012 IEEE International Conference on Acoustics, Speech and Signal Processing (ICASSP)* (pp. 5349-5352). IEEE.

Kepner, J., Arcand, W., Bestor, D., Bergeron, B., Byun, C., Gadepally, V., et al. (2014). Achieving 100,000,000 database inserts per second using Accumulo and D4M. *IEEE High Performance Extreme Computing (HPEC)*. IEEE.

Samsi, S., Brattain, L., Arcand, W., Bestor, D., Bergeron, B., Byun, C., et al. (2016). Benchmarking SciDB Data Import on HPC Systems. *IEEE High Performance Extreem Computing (HPEC)*. IEEE.

Stonebraker, M., Brown, P., Poliakov, A., & Raman, S. (2011). The Architecture of SciDB. *International Conference on Scientific and Statistical Database Management*. Springer.

Stonebraker, M., Brown, P., Poliakov, A., & Raman, S. (2011). The Architecture of SciDB. *International Conference on Scientific and Statistical Database Management*. Springer.

Weale, T., Gadepally, V., Hutchison, D., & Kepner, J. (2016). Benchmarking the Graphulo Processing Framework. *IEEE High Performance Extreme Computing (HPEC)*. IEEE.